\documentclass[10pt,final,journal]{IEEEtran}
\usepackage[utf8]{inputenc}

\IEEEoverridecommandlockouts
\usepackage{xcolor}
\usepackage{amsmath,amssymb,amsfonts}
\usepackage{mathtools}
\usepackage{array}
\usepackage{tikz}
\usepackage{cite}
\usepackage{textcomp}
\usepackage{gensymb}
\usepackage{subfigure}
\usepackage{graphicx}
\usepackage{bm}
\usepackage{mathrsfs}
\usepackage[ruled]{algorithm2e}
\usepackage{url}
\usepackage{multirow}

\begin{document}

\title{Topology Design for GNSSs Considering Both Inter-satellite Links and Ground-satellite Links}

\author{
Zhibo Yan, 
Kanglian Zhao,~\IEEEmembership{Member,~IEEE,}
Wenfeng Li,~\IEEEmembership{Member,~IEEE,}\\
Chengbin Kang,
Jinjun Zheng,
Hong Yang 
and Sidan Du,~\IEEEmembership{Member,~IEEE}
\thanks{
 
\textit{(Corresponding author: Kanglian Zhao.)}
}
\thanks{Z. Yan and S. Du are with the School of Electronic Science and Engineering, Nanjing University, Nanjing 210046, China (e-mail: zhiboyan\_nju@smail.nju.edu.cn; coff128@nju.edu.cn).}
\thanks{K. Zhao and W. Li are with the School of Electronic Science and Engineering, Nanjing University, Nanjing 210046, China, and also with the Peng Cheng Laboratory, Schenzhen 518052, China (e-mail: zhaokanglian@nju.edu.cn; leewf\_cn@hotmail.com).}
\thanks{C. Kang, J. Zheng, and H. Yang are with the Institute of Spacecraft System Engineering, China Academy of Space Technology (CAST), Beijing 100094, China (e-mail: 75012565@qq.com, zhjinjun@vip.sina.com; 13381105509@189.cn).}
}

\maketitle

\begin{abstract}
   Inter-satellite links (ISLs) are adopted in global navigation satellite systems (GNSSs) for high-precision orbit determination and space-based end-to-end telemetry telecommand control and communications. 
   Due to limited onboad ISL terminals, the polling time division duplex (PTDD) mechanism is usually proposed for space link layer networking.
   By extending the polling mechanism to ground-satellite links (GSLs), a unified management system of the space segment and the ground segment can be realized.
   However, under the polling system how to jointly design the topology of ISLs and GSLs during every slot to improve data interaction has not been studied.
   In this paper, we formulate the topology design problem as an integer linear programming, aiming at minimizing the average delay of data delivery from satellites to ground stations while satisfying the ranging requirement for the orbit determination.
   To tackle the computational complexity problem, we first present a novel modeling method of delay to reduce the number of decision variables. 
   Further, we propose a more efficient heuristic based on maximum weight matching algorithms. 
   Simulation results demonstrate the feasibility of the proposed methods for practical operation in GNSSs.
   Comparing the two methods, the heuristic can achieve similar performance with respect to average delay but with significantly less complexity.
   
\end{abstract}


\section{Introduction}
Global navigation satellite systems (GNSSs) play a pivotal role in providing ubiquitous, continuous and reliable positioning, navigation, and timing (PNT) services \cite{PNT}. 
Recently, inter-satellite links (ISLs) are being introduced into GNSSs to reduce dependence on ground stations (GSs) as GSs are able to control satellites out of sight through ISLs.
Radiofrequency narrow-beam antennas such as phased array antennas are one of the feasible solutions for deploying ISLs on satellites.
While the narrow-beam antennas bring the advantages of strong anti-jamming capability, high ranging accuracy, low power consumption, and high data rate\cite{yang2017}, their directionality plus platform restrictions result in the limited number of ISLs that a satellites can establish simultaneously.

Typically a navigation satellite is equipped with one narrow-beam antenna because of satellite platform restrictions, which renders intermittently connected topology in GNSSs \cite{huang2018,Sun2018}.
That is, the whole space network is composed of pairs of nodes at any moment, and there does not exist end-to-end paths between two arbitrary nodes.
Thus to enhance data communications and orbit determination, the polling time division duplex (PTDD) ranging hierarchy of the ISLs is adopted by GPS\cite{gps}, Galileo GNSS+\cite{Galilo}, and Beidou\cite{beidou}.
In the polling mechanism, satellites take time slots as the basic unit to communicate with different satellites in different slots. 
During a specific slot, multi-point to multi-point communication links are established and the store-carry-and-forward method\cite{yang2018} is adopted to realize end-to-end data transmission.

In ISL-enabled GNSSs, high-precise orbit determination necessitates each satellite measuring enough distinct satellites, i.e., setting up sufficient ISLs with different counterparts\cite{yangdaoning2017}, which is the so-called ranging requirement. 
On the other hand, the ground monitoring station of GNSSs needs to gather telemetry data from navigation satellites and send telecommand data to them.
These delay-sensitive data require the satellites that are out of sight of GSs to frequently establish ISLs with satellites that are visible to GSs.
Besides, emerging additional applications like global short messages communication (GSMC) in Beidou \cite{yang2019,LI2020} should also provide services with low delay.
Thus how to schedule the topology taking into consideration both the ranging and the communication requirements, i.e, the topology design \cite{JuanSurvey,Juan2014} determines the performance of GNSSs.
Different from traditional terrestrial networks, satellite networks are characterized by sporadic but predictable visibility caused by satellite movements, which poses another challenge to this problem.

Up to now, many recent works have designed several algorithms based on heuristics \cite{Yan2015,huang2018,Sun2018,Hou2018}, and the authors have also proposed some methods to solve the topology design problem \cite{YanWCL,YanTAES}.
Nevertheless, all these surveyed works assume that the ground-satellite links (GSLs) are continuous connection such that data on satellites visible to GSs can be transferred to GSs immediately regardless of the forwarding queue.
Thus these works focus on the delay from satellites invisible to GSs to satellites visible to GSs, and neglect the delay from the visible satellites to GSs.
However, the incompatibility of the time-slotted system of ISLs and the continuous connection of GSLs would render inefficient data interaction.
To address this issue, the ground segment could adopt the same PTDD system as the space segment by deploying GSs equipped with narrow-beam antennas.
Before the ability of real autonomous navigation is realized, GSLs could share the same communication and measurement system with ISLs\cite{Ren2019,Bai2020}, in other words, a unified management of the space segment and the ground segment can be achieved.
For the satellites visible to GSs, whether to set up ISLs with other satellites or establish GSLs with GSs at a specific time slot, has a great impact on the ranging and communication performance of the GNSS network.
This problem, however, has not received enough attention.

In order to fill this gap, by jointly considering the ISLs and the GSLs, we study the topology design for GNSSs where the GSLs adopt PTDD like the ISLs in this paper. 
To the best of our knowledge, our work is the first in proposing and solving such a problem.
In more detail, the contributions of this paper are highlighted as follows:
\begin{itemize}
    \item We exploit the finite state automaton (FSA) \cite{Chang1998} model to capture the dynamic visibility of satellite networks and formulate the topology design problem in GNSSs as an integer linear programming (ILP) to minimize the overall delay of all data packets from satellites to ground stations.
    \item The time complexity of the ILP rises exponentially with the increase of the network size and the number of time slots in each state. By introducing a novel modeling of network delay in GNSSs, we first propose another method also based on ILP but with much fewer integer decision variables, i.e., a variant of ILP. However, due to the complexity of ILP by nature, the proposed method still faces the problem of not being able to schedule rapidly. Thus we further design an efficient heuristic based on maximum weight matching to solve the topology design problem.
    \item Extensive simulations are conducted to evaluate the performance of our proposed algorithms. 
    Both the proposed two methods are able to solve the topology design problem for practical operation in GNSSs with large scale. 
    Furthermore, the heuristic could achieve performance similar to the ILP variant but with significantly less computational complexity.
    Finally, the performance of the proposed methods are further evaluated with respect to several critical network parameters.   
\end{itemize}

The rest of the paper is structured as follows. 
We briefly review the current topology design solutions both for GNSSs and other satellite networks in Section II. 
Section III introduces the system model and elaborates the ILP method for the network delay optimization. 
In Section IV, we propose a novel modeling of communication delay with fewer decision variables to solve the topology design problem.
We further propose a heuristic with lower time complexity in Section V.
Simulation results are presented and analyzed in Section VI.
The paper is finally concluded with Section VII.
\section{Related Work}
\subsection{Topology Design in Remote Sensing and Communication Satellite Networks}
Topology design, scheduling which possible links to set up, determines the overall performance of the network, and has been studied in remote sensing satellite networks \cite{Juan2014, Juan2016, Zhou2017, Zhou2019} and in the emerging mega-constellation for satellite communication \cite{Bhattacherjee2019}.
In \cite{Juan2014}, Juan \textit{et al.} proposed a fair contact plan to maximize the fairness of the overall topology without impairing the network capacity. 
Besides, the authors also proposed to solve the topology design problem as a matching problem to achieve polynomial time complexity.
In \cite{Juan2016}, the authors took the traffic into consideration and modeled the topology design as a mixed integer linear programming.
In \cite{Zhou2017}, Zhou \textit{et al.} introduced a more complex scenario and formulated the corresponding constraints into MILP, e.g., the differentiation for missions and practical energy harvesting.
The authors then proposed a primal decomposition based method to decompose the MILP model and also devised a heuristic on a slot-by-slot basis.
In order to cover the unknown traffic in the future, the authors went on to study the robust planning without full distribution information of the long-term data arrival in \cite{Zhou2019}. 
With the emergence of mega-constellations, how to design the topology for inter-satellite networks is drawing attention from industry and academia.
Due to the massive number of satellite nodes, traditional methods for topology design like linear programming become inefficient even ineffective. 
Thus in \cite{Bhattacherjee2019}, Bhattacherjee \textit{et al.} exploited a repetitive pattern called \textit{motif} in the network topology to optimize the scalability of the proposed topology design algorithm, while still providing near-minimal network latencies compared with the past methods. 
\subsection{Topology Design in GNSSs}
In the context of GNSSs, the topology design problem is characterized by the need to consider the ranging performance.
In \cite{Yan2015}, Yan \textit{et al.} first introduced FSA into GNSSs and optimized the delay of the network by means of the simulated annealing algorithm, in which the branch and exchange method guaranteed the satisfaction of the ranging requirement when generating new solutions. 
In \cite{huang2018}, Huang \textit{et al.} proposed a cascade optimization design to optimize the topology and the parameters used in the FSA model. 
In \cite{Sun2018}, a generic algorithm was used to find out the best routes for satellites, then these routes were extended and supplemented with ranging links to meet the ranging requirement.
Since the topology design in GNSSs improves the ranging performance at the cost of communication delay, Hou \textit{et al.} studied the trade-off between inter-satellite communication and ranging in \cite{Hou2018}.
In our previous work \cite{YanWCL}, we focused on the delay of telemetry and for the first time formulated the topology design in GNSSs as an ILP.
The proposed method could meet the ranging requirement at the lowest cost of link resources, and thus the communication performance was optimized.
Besides, we also investigated the possibility of distributed topology control of GNSSs in \cite{YanTAES}, where the requirements were discussed and a suitable method was proposed to achieve real autonomy. 
However, all these surveyed papers assume that data would sink at satellites visible to GSs and neglect the data flows between the satellites and GSs, which is not suitable if the connection of GSLs is not continuous.
The study of topology design in GNSSs considering the time-slotted GSLs is still missing in the literature.
Thus in this paper, we concentrate on designing efficient topology design methods for GNSSs where the GSLs still adopt the PTDD hierarchy.

\section{System Model and Problem Formulation}
\label{sec: system model}

\subsection{FSA Model and PTDD Ranging Hierarchy}
\begin{figure}[tp]
    \centering
    \includegraphics[width=0.48\textwidth]{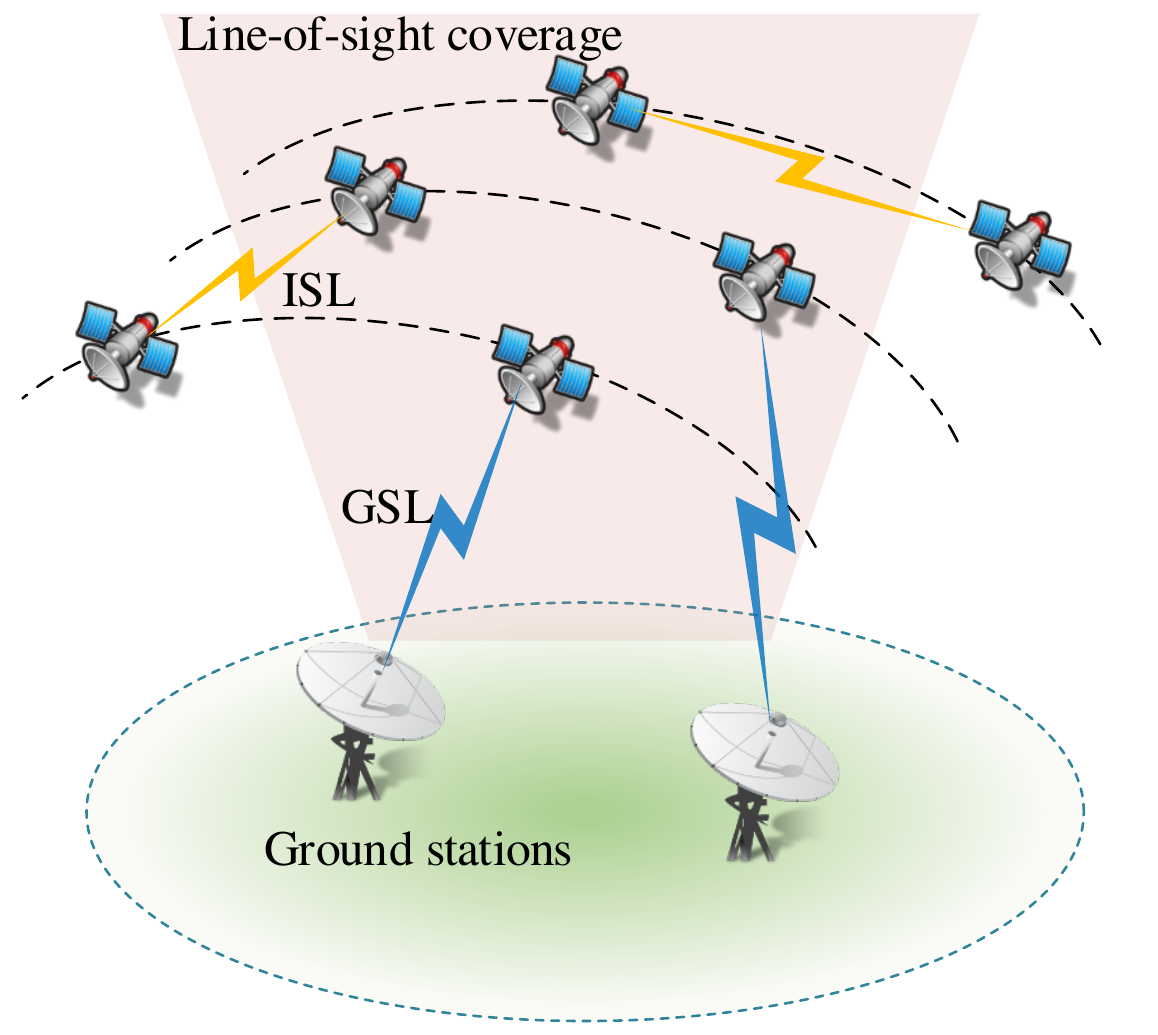}
    \caption{Topology in GNSS at a given time.}
    \label{fig: GSL}
\end{figure}

In addition to the challenges faced by traditional terrestrial networks, another non-trivial problem that needs to be tackled in satellite networks is the dynamics caused by satellite movements. 
As satellites orbit the earth, the ISLs between the satellites and the GSLs between satellites and GSs are unstable and hereby change over time.
In order to circumvent such a problem, FSA has been widely adopted by researchers since it was proposed in \cite{Chang1998}.
FSA divides the whole schedule period into a series of equal-length time intervals, which are called states and can be denoted by $s \in {\cal S} = \{1,\, 2 \, ...\, S\}$.
In each state, two entities (including satellites and GSs) are defined as visible iff they are visible to each other throughout the state.
Then the visibility of each pair of satellites and between satellites and GSs is regarded as static in each state, such that topology design can be solved on the basis of fixed visibility.
Furthermore in the context of GNSSs, improving the precision of autonomous navigation necessitates each satellite frequently switching ISLs to different satellites within a relatively short time to gain more ranging information \cite{Yan2015}.
To this end, a time-slotted system is adopted in GNSSs where each FAS state $s \in {\cal S}$ is further divided into several equal-length time slots, complying with the PTDD hierarchy.
We index the slots in one state by $t \in {\cal T} = \{1,\, 2 \, ...\, T\}$. 
Satellites are then able to set up ISLs with different satellites in various time slots.

\subsection{Basic Constraints}
We consider a general GNSS as shown in Fig. \ref{fig: GSL}, which consists of $N^s$ satellite nodes denoted by ${\cal V}^s$ and $N^g$ ground station nodes denoted by ${\cal V}^g$. 
We denote all the nodes in such a network as ${\cal V} = {\cal V}^s \cup {\cal V}^g = \{v_i | 1 \le i \le N\}$, where $N$ is the total number of nodes and $N = N^s + N^g$.
Then the set of satellite nodes is further divided into non-anchor satellites and anchor satellites, expressed as  ${{\cal V}^s} = {\cal V}^n \cup {\cal V}^a$.
Anchor satellites are defined as those who are visible to any GS (i.e., within the line-of-sight coverage of any GS) in the considered state, and on the contrary non-anchor satellites are defined as those who are invisible to all GSs (i.e., beyond the line-of-sight coverage of all GSs).
Let $N^n$, $N^a$ denote the number of non-anchor satellites, anchor satellites respectively, we have $N = N^n + N^a + N^g$.
In this paper, we take one state as an example to show how the constrains are formulated and how the topology design problem is solved. 
For a given state $s\in \cal S$, we introduce two 0-1 matrices $\bm X$ and $\bm Y$ to denote the designed topology and the visibility, respectively. 
Wherein, $x_{i,j,t}=1$ represents there is a link scheduled between $v_i$ and $v_j$ in the $t$-th time slot and 0 otherwise.
In $\bm Y$, $y_{i,j}=1$ indicates $v_i$ and $v_j$ are visible to each other, and 0 otherwise.
The constraints between $\bm X$ and $\bm Y$ are given by
\begin{equation}
    x_{i,j,t}=\{0,1\}, 
    \quad \forall v_i,v_j\in {\cal V}, t\in {\cal T},
\end{equation}
\begin{equation}
\label{eqn: bi}
    x_{i,j,t}=x_{j,i,t}, 
    \quad \forall v_i,v_j\in {\cal V},t\in {\cal T},
\end{equation}
and
\begin{equation}
\label{eqn: visi}
    x_{i,j,t} \le y_{i,j},
    \quad \forall v_i,v_j\in {\cal V}, t\in {\cal T}.
\end{equation}
Wherein, (2) restricts that the links are bi-directional, and (3) states that a link can be assigned between $v_i$ and $v_j$ only when they are visible to each other.
Besides in GNSSs, due to limited terminals onboard, each node can only establish one link during a time slot.
That is, a satellite node will communicate either with another satellite or with a ground station in a time slot.
Thus we have
\begin{equation}
\label{eqn: basic}
    \sum\limits_{v_j\in {\cal V}} {x_{i,j,t}}  \le 1,  
    \quad \forall v_i\in {\cal V}, t \in {\cal T}.
\end{equation}

\subsection{Ranging Constraints}
In accordance with the PTDD ranging hierarchy, a satellite needs to complete multiple times of ranging with other satellite nodes.
Given the minimum required number of ranging links $L^{\min}$, the ranging constraints can be expressed by
\begin{equation}
\label{eqn: ranging links} 
{\ell _{i,j}} =
\begin{cases}
    {1,} & {\sum\limits_{t \in {\cal T}} {x_{i,j,t} \ge 1}} \\
    {0,} & {\sum\limits_{t \in {\cal T}} {x_{i,j,t} = 0} }
\end{cases}
    \quad \forall v_i,v_j \in {{\cal V}^s},
\end{equation}
\begin{equation}
    \sum\limits_{v_j \in {\cal V}^s} {{\ell_{i,j}} \ge {L^{\min }}, 
    \quad \forall v_i\in {\cal V}^s}.
\end{equation}
Here, $\ell_{i,j}$ describes whether satellite nodes $v_i$ and $v_j$ have established an ISL with each other in $T$ time slots. 
Note that repeated assignments of a link between two satellites in several time slots do not contribute to the ranging performance.
However, (5) is not a linear expression, thus the following modeling method is used to transform (5) to a linear one:
\begin{equation}
    {{\ell}_{i,j}} = \{ 0, 1\} ,  
    \quad \forall v_i,v_j\in {{\cal V}^s},
\end{equation}
\begin{equation}
\label{eqn: M}
    {\ell _{i,j}} \le \sum\limits_{t \in {\cal T}} {x_{i,j,t} \le {\dot M} \, { \ell _{i,j}}} , \quad \forall v_i,v_j\in {{\cal V}^s}.
\end{equation}
Wherein ${\dot M} $ is a constant greater than the sum of $x_{i,j,t}$.
If $\sum\nolimits_{t \in {\cal T}} {x_{i,j,t}} =0$, to satisfy ${\ell _{i,j}} \le \sum\nolimits_{t \in {\cal T}} {x_{i,j,t}}$, we have ${\ell _{i,j}}=0$.
On the other hand, if $\sum\nolimits_{t \in {\cal T}} {x_{i,j,t}} \ge 1$, to satisfy $ \sum\nolimits_{t \in {\cal T}} {x_{i,j,t}} \le {\dot M} \, { \ell _{i,j}}$, ${\ell _{i,j}}=1$ will hold.

\subsection{Flow Conservation Constraints}
In GNSSs, all satellite nodes would send telemetry data back to ground stations and part of the satellites are assumed to provide other services like GSMC in Beidou for users.
We denote the total data of all services for all nodes along all slots by using a matrix ${\bm F}_{{N^s}\times{T}}=\{ f_{i,t}|\forall v_i \in {\cal V}^s, t\in {\cal T} \}$, where $f_{i,t}$ represents the data that satellite $v_i$ generated in the $t$-th time slot.   
Let $f^{td}$ denote the set of telemetry data (in packets) that each satellite generated in each slot.
For simplicity, we assume the total data of other services in each slot are also fixed and is given by $f^{sm}$.
That is, for some satellites $f_{i,t}=f^{td}$, and for the other satellites who are able to provide special services $f_{i,t}$ is given by the sum of $f^{td}$ and $f^{sm}$.
For each $f\in \bm F$, we use a integer matrix ${\bm R}^f$ to denote the route, and the integer variable $r^f_{i,j,t}$ in ${\bm R}^f$ denotes the amount of traffic $f$ forwarded from $v_i$ to $v_j$ in the $t$-th time slot. 
Besides, we define $b_{i,t}^f$ (in packets) as the volume of data for traffic flow $f$ in the buffer of satellite $v_i$ at the end of the $t$-th time slot.
In a manner akin to early work \cite{Zhou2017, Juan2016} on remote sensing satellite networks, we model flow conservation constraints here.
For the nodes as the data sources, the flow conservation is described as the sum of all outgoing flows and data in the buffer equals the amount of generated data, that is
\begin{equation}
   \sum\limits_{v_j\in {\cal V}} r_{i,j,t}^f +  b_{i,t}^f = f_{i,t},      
 \, v_i = \textit{Src}(f), t = \textit{St}(f), \forall f \in \bm F,
\end{equation}
where $\textit{Src}(f)$ is the source satellite of $f$ and $\textit{St}(f)$ is the staring time slot of $f$.
In specific, $\textit{Src}(f_{i,t})=v_i$ and $\textit{St}(f_{i,t})=t$.
For the satellites that act as relay nodes, we have 
\begin{equation}
\begin{aligned}
    \sum\limits_{v_j \in {\cal V}} r_{i,j,t}^f +  b_{i,t}^f = & \sum\limits_{v_j\in {\cal V}} r_{j,i,t}^f + b_{i,t-1}^f,\\
   & v_i\in {\cal V}^s, v_i \neq \textit{Src}(f), t\neq \textit{St}(f), \forall f \in \bm F.
\end{aligned}
\end{equation}
We set an upper bound $B^{\max}$ for each buffer on satellites, thus
\begin{equation}
    \sum\limits_{f\in {\bm F}} b_{i,t}^f \le B^{\max},
    \forall v_i \in {\cal V}^s, t\in {\cal T}.
\end{equation}
Furthermore, the aggregated data route from $v_i$ to $v_j$ for all $f$ cannot exceed the capacity of the corresponding link, thus we have 
\begin{equation}
\label{eqn: css}
    \sum\limits_{f\in {\bm F}} r_{i,j,t}^f \le Css,
    \quad  \forall v_i \in {\cal V}^n, v_j \in {\cal V}^a,  t\in {\cal T},
\end{equation}
\begin{equation}
\label{eqn: csg}
    \sum\limits_{f\in {\bm F}} r_{i,j,t}^f \le Csg,
    \quad \forall v_i \in {\cal V}^a, v_j \in {\cal V}^g,   t\in {\cal T},
\end{equation}
\begin{equation}
    \sum\limits_{f\in {\bm F}} r_{i,j,t}^f = 0,
    \quad \forall v_i \in {\cal V}^n, v_j \in {\cal V}^n,   t\in {\cal T},
\end{equation}
\begin{equation}
    \sum\limits_{f\in {\bm F}} r_{i,j,t}^f = 0,
    \quad \forall v_i \in {\cal V}^a, v_j \in {\cal V}^a,   t\in {\cal T},
\end{equation}
where $C_{ss}$ and $C_{sg}$ are the capacity (in packets/slot) of ISLs and GSLs respectively.
For the sake of simple routing, we assume non-anchor satellites can only directly send data to an anchor satellite, otherwise the data should be stored and wait until an earliest connection with an anchor satellite \cite{YanWCL}.
The same is true between anchor satellites and GSs.
A specific link will not be able to carry traffic unless enabled.
To achieve this a sufficiently big coefficient $M$ is introduced
\begin{equation}
\label{eqn: ILP last}
     \sum\limits_{f\in {\bm F}} r_{i,j,t}^f \le M\, x_{i,j,t},
     \quad \forall v_i, v_j \in {\cal V}, t\in {\cal T}.
\end{equation}
When the link between $v_i$ and $v_j$ is inactive in some slot, i.e., $x_{i,j,t}=0$, the right part of (\ref{eqn: ILP last}) is zero, rendering $\sum\nolimits_{f\in {\bm F}} r_{i,j,t}^f$ equal to zero as well.
When $x_{i,j,t}=1$, flows could pass through $v_i$ to $v_j$, i.e., the active link.
Thus in order to agree with (\ref{eqn: css}) and (\ref{eqn: csg}), $M$ must take the value greater than both $C_{ss}$ and $C_{sg}$.
\subsection{Problem Formulation}
Up to now we have formulated all the constraints in GNSSs.
Our objective is to minimize the delay to GSs of all flows in the traffic profile $\bm F$.
To achieve this optimization, we formulate the topology design problem in GNSSs as follows: 
\begin{equation}
\label{eqn:obj}
\begin{split}
\mathop{\min}\limits_{\bm{X}}\quad & \sum\limits_{f\in {\bm F}} \sum\limits_{v_i\in {{\cal V}^s}} \sum\limits_{t\in {\cal T}} (t-\textit{St}(f)) \, b_{i,t}^f\\
\text{ s.t. }\quad & (1) \textit{-}(4), (6)\textit{-}(16),\\ 
\end{split}
\end{equation}
which falls into the category of ILP that is proved to be NP-hard.
Recall that $\textit{St}(f)$ is the staring time slot of $f$. 
As the weight factor $(t-\textit{St}(f))$ grows with the $t$, the optimization forces each flow generated on satellites to be relayed to GSs as soon as possible.
Otherwise the residual data would occupy the buffer on satellites and thus increase the objective function.
During the same $t$-th slot, data generated earlier (i.e., smaller $\textit{St}(f)$) would be transferred with higher priority because of higher weight factor (i.e., bigger $(t-\textit{St}(f))$). 
The time complexity of ILP is dominated by the number of variables and constraints\cite{Bertsekas1997}, which renders limited scalability of the method in practical scenarios.
On our computing platform, ILP fails to finish within a reasonable time, i.e., 2 days for practical GNSSs. 
The contributing factor is the high-dimensional decision variable matrix $\bm R$.
Since $\bm F$ is $N\times K$, $\bm R$ is a five-dimensional matrix, i.e., $N \times N \times K \times N \times K$.
In our simulation, when $N$ takes the value of 30 and $K$ takes the value of 20\cite{yang2017,Gao2012}, $\bm R$ will be in the order of tens of millions, making the optimization unsolvable even by supercomputers because the amount of computation increases exponentially with the increase in the number of decision variables. 
Thus in the next sections, we focus on devising sub-optimal methods that scale better compared with the ILP model in this section.

\section{Routing-Agnostic Network Delay Minimization}
\label{sec: RAILP}

\begin{figure*}[htp]
    \centering
    \includegraphics[width=0.98\textwidth]{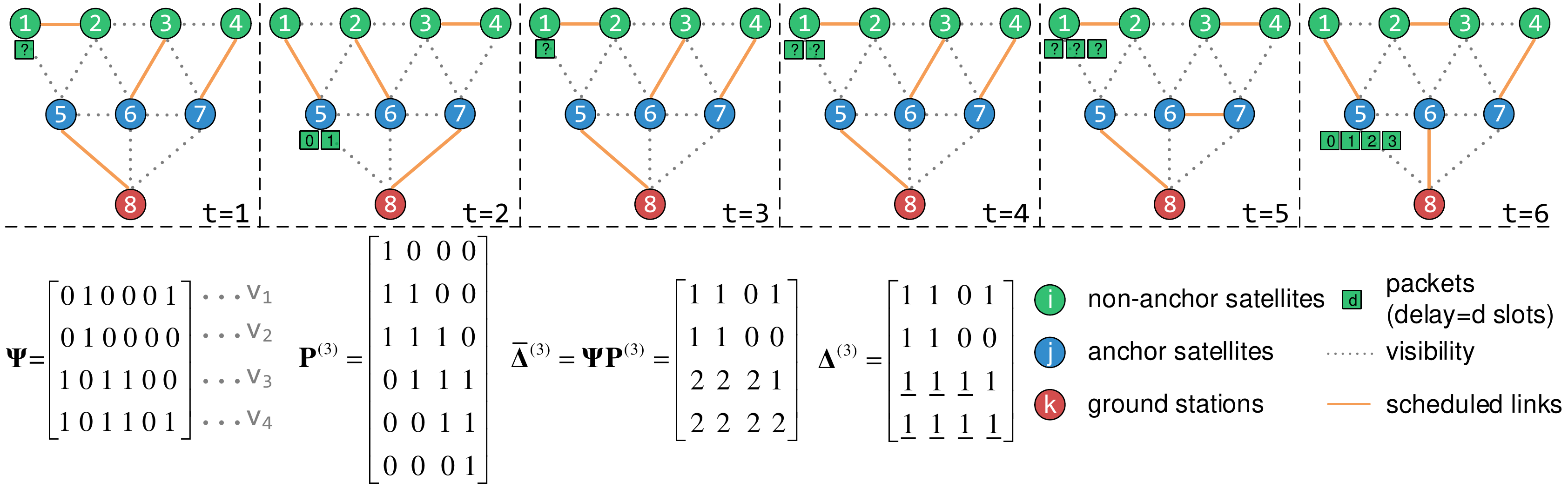}
    \caption{Modeling of communication delay for non-anchor satellites to anchor satellites. The network is composed of 8 nodes with $T=6$. For simplicity, only the delay of data generated by each time slot on satellite $v_1$ is showed and the probe matrix ${\bm P}^{(1)}$ is given as an example. Probe matrices ${\bm P}^{(1)}$, ${\bm P}^{(2)}, {\bm P}^{(4)}, {\bm P}^{(5)}$, and ${\bm P}^{(6)}$ and corresponding other matrices are omitted. 
    }
    \label{fig: delay}
\end{figure*}

Since the immense number of decision variables of the routing matrix $\bm R$ lead to the infeasible ILP model in practical GNSSs, we propose a new modeling of communication delay with fewer variables to solve the proposed topology design problem. 
The proposed method proactively disregards the route scheduling of each flow $f$, thus is called routing-agnostic ILP (RAILP) in this paper. 

\subsection{Modeling of Communication Delay}
In this section, we assume once an anchor satellite establishes a GSL with a GS, the former could send all data to the latter.
The same is true when a non-anchor satellite is connected to any anchor satellite.
Thus for anchor satellites, the communication delay of data can be described as the waiting time slots before being scheduled with a GSL.  
As long as anchor satellites establish GSLs with GSs more frequently, data would be transferred to GSs with lower delay.
In the same spirit, non-anchor satellites desire to establish ISLs with anchor satellites as much as possible as data need to be first forwarded to anchor satellites, then relayed to GSs.
To simplify the ILP model above, we ignore the effects of queue delay, buffer limitation, \textit{etc.}, and model the communication delay merely based on the frequency of link establishment, including GSLs and ISLs between non-anchor satellites and anchor satellites, which support the data transmission.

For the delay from non-anchor satellites to anchor ones, we introduce a 2-dimensional matrix $\bm{\psi}_{{N^n}\times T}$ to denote whether a non-anchor satellite is connected to an anchor satellite. 
The element $\psi _{i,t}$ in $\bm{\Psi}$ equals 1 means the $i$-th non-anchor satellite has an ISL with any anchor satellite in the $t$-th time slot, which can be expressed as
\begin{equation}
\label{}
    {\psi _{i,t}} = \sum\limits_{v_j \in {{\cal V}^a}} {x _{i,j,t}},   
    \quad \forall v_i \in {{\cal V}^n},t\in {\cal T}.
\end{equation}
As illustrated in Fig. \ref{fig: delay}, non-anchor satellite $v_1$ has access to anchor satellites in two out of six time slots (the 2nd and the 6th time slots). 
Thus the first row of $\bm{\Psi}$ is $\left[0\;1\;0\;0\;0\;1 \right]$.
Since the non-anchor satellites are able to send data only when they are connected to anchor satellites, the delay of data generated in all 6 slots are $\left[1\;0\;3\;2\;1\;0 \right]$, which is depicted in the figure.
It is worth noting that the maximum delay of one non-anchor satellite is indeed the maximum number of consecutive zeros in the corresponding row in $\bm{\Psi}$.
Assuming there are $k$ consecutive zeros, the overall delay of data generated in these time slots is $k+(k-1)+...+1$. 

In order to model the exact delay to anchor satellites of each non-anchor satellite, we construct $T$ probe matrices from ${\bm P}^{(1)}$ to ${\bm P}^{(T)}$.
The $t$-th probe matrix can be defined as
\begin{equation}
     {\bm{P}}^{(t)}_{{T}\times (T-t+1)} = ({ \bm{P}}^{(t)}_1, \, { \bm{P}}^{(t)}_2 \, ...\, { \bm P}^{(t)}_{T-t+1}),
     \quad \forall t \in \cal T,
\end{equation}
where $ { \bm P}^{(t)}_{i}$ is the $i$-th column of ${ \bm P}^{(t)}_{{T}\times (T-t+1)}$, and
\begin{equation}
      { \bm P}^{(t)}_{i} = (p_{1,i} \, ...\, p_{i-1,i},\, p_{i,i}\,...\,p_{i+t-1,i},\,p_{i+t,i}\,...\,p_{T,i} )'.
\end{equation}
It can be noted that ${ \bm P}^{(t)}_{i}$ is composed of three parts.
$p_{1,i}$ to $p_{i-1,i}$ take the value of 0, $p_{i,i}$ to $p_{i+t-1,i}$ are 1, and $p_{i+t,i}$ to $p_{T,i}$ take the value of 0.
Note that the first part does not exist in the first column ${\bm P}_1^{(t)}$ (i.e., $i=1$) and
 the third part does not exist in the last column ${\bm P}_{T-t+1}^{(t)}$  (i.e., $i=T-t+1$).
Besides, when $t=T$, there is only one column in ${\bm P}^{(t)}_{{T}\times 1}$ and all elements are 1.
In general, as a probe matrix, ${ \bm P}^{(t)}$ is able to detect how many successive $t$ zeros are in each row of $\bm{\Psi}$, i.e., for each non-anchor satellite.
In specific, we could obtain such a matrix as
\begin{equation}
    \bar{\bm{ \Delta}} ^{t} = \bm{{\Psi}}\,  {\bm P}^{(t)},
    \quad \forall t \in \cal T
\end{equation}
where each 0 in $\bar{\bm{ \Delta}} ^{{(t)}}$ represents there are successive $t$ zeros in the corresponding place in  $\bm{\Psi}$.
For instance in Fig. \ref{fig: delay}, ${ \bm P}^{(3)}$ is used for detecting three consecutive zeros in $\bm{\Psi}$.
From the $\bar{\bm{ \Delta}} ^{{(3)}}$, we can see $\bar{\bm{ \Delta}} ^{(3)}_{1,3}$, $\bar{\bm{ \Delta}} ^{(3)}_{2,3}$, and $\bar{\bm{ \Delta}} ^{(3)}_{2,4}$ are 0, which is consistent with the number of consecutive three zeros in $\bm{\Psi}$.
It should be noted that  we could extract two consecutive three zeros, i.e., $\left[ 0\;0\;0\right]$, from four consecutive zeros.
By multiplying with ${ \bm P}^{(1)}$ to ${ \bm P}^{(T)}$, we could get the situation of different number of consecutive zeros in $\bm {\Psi}$. 
Thus for the three consecutive zeros of $v_1$ from the 3-th to the 5-th time slots, we could get three 0 in the first row of $\bar{\bm{ \Delta}}^{(1)}$ (i.e., three $\left[ 0\right]$), two 0 in the first row of $\bar{\bm{ \Delta}}^{(2)}$ (i.e., two $\left[ 0\;0\right]$), and one 0 in the first row of $\bar{\bm{\Delta}}^{(3)}$ (i.e., one $\left[ 0\;0\;0\right]$). 
Thus we can get the conclusion that, if there are $k$ consecutive zeros in some row of $\bm{\Psi}$, accordingly there will be $k+(k-1)+...+1$ zeros in $\bar{\bm{ \Delta}} ^{(1)}$ to $\bar{\bm{ \Delta}} ^{(T)}$.
Recall that $k+(k-1)+...+1$ is exactly the delay we obtained for a non-anchor satellite that does not access to any anchor satellite for $k$ consecutive time slots.
Thus the total number of zeros in $\bar{\bm{ \Delta}} ^{(1)}$ to $\bar{\bm{ \Delta}} ^{(T)}$ is the delay for all non-anchor satellites to anchor satellites along $T$ time slots.

Given the number of time slots $T$ and the number of non-anchor satellites $N^n$, the total number of elements in $\bar{\bm{ \Delta}} ^{(1)}$ to $\bar{\bm{ \Delta}} ^{(T)}$ is certain.
If we maximize the number of non-zeros in $\bar{\bm{ \Delta}} ^{(1)}$ to $\bar{\bm{ \Delta}} ^{(T)}$, the number of zeros are minimized, and also the overall delay.  
To this end, the matrix ${\bm{ \Delta}} ^{(t)}$ has come to be used to convert a number greater than 1 to 1 in a manner akin to (\ref{eqn: M}).
We have
\begin{equation}
    \bm{{\Delta}} ^{(t)} \le {\bar {\bm \Delta}} ^{(t)} \le {\bar M} \, {\bm \Delta} ^{(t)},
    \quad \forall t \in {\cal T}, 
\end{equation}
and each element in $\bm{{\Delta}} ^{(t)}$ is a binary variable. 
Then the sum of all elements (i.e., 1) in ${\bm \Delta} ^{(t)}$ is the number of non-zeros in $\bar {\bm \Delta} ^{(t)}$.
The differences between $\bar{\bm{ \Delta}} ^{(t)}$ and $\bm{{\Delta}} ^{(t)}$ are underlined as shown in Fig.  \ref{fig: delay}.
Thus our objective is to maximize
\begin{equation}
    {{\Gamma}^{n \to a}} =\sum\limits_{i}{f_i} \sum\limits_{t\in \cal T} \sum\limits_j {\bm \Delta}_{i,j} ^{(t)},
\end{equation}
where $\sum\limits_{t\in \cal T} \sum\limits_j {\bm \Delta}_{i,j} ^{(t)}$ is the total number of non-zeros for the $i$-th non-anchor satellite, and $f_i$ is the corresponding weight factor obtained from traffic flows.
Since we assume that the traffic generated by all satellites in each time slot is fixed, $f_i$ can be simply expressed as $f_{i,1}$.

As for the delay from anchor satellites to GSs, we also use the same modeling method which can be expressed as (\ref{eqn: a-g 1})-(\ref{eqn: a-g 2}). We have
\begin{equation}
\label{eqn: a-g 1}
    {\phi _{i,t}} = \sum\limits_{v_j \in {{\cal V}^g}} {x _{i,j,t}},   
    \quad \forall v_i \in {{\cal V}^a},t\in {\cal T},
\end{equation}
\begin{equation}
    {\tilde {\bm \Lambda}} ^{(t)} = {\bm{\Phi}}\,  {\bm P}^{(t)},
    \quad \forall t \in \cal T,
\end{equation}
\begin{equation}
    {\bm \Lambda} ^{(t)} \le {\tilde {\bm \Lambda}} ^{(t)} \le {\tilde M} \, {\bm \Lambda} ^{(t)},
    \quad \forall t \in {\cal T},
\end{equation}
\begin{equation}
\label{eqn: a-g 2}
    {{\Gamma}^{a \to g}} =\sum\limits_{i}{f_i} \sum\limits_{t\in \cal T} \sum\limits_j {\bm \Lambda}_{i,j} ^{(t)},
\end{equation}
where $\phi _{i,t}$ in $\bm{\Phi}$ equals 1 means the $i$-th anchor satellite establishes a GSL with some GS in the $t$-th time slot, and 0 otherwise. 
$\bar{\bm{ \Lambda}} ^{(t)}$ and ${\bm{ \Lambda}} ^{(t)}$ are used for detecting successive $t$ zeros in $\bm{\Phi}$, and ${{\Gamma}^{a \to g}}$ is part of our optimization objective.

\subsection{The Proposed RAILP}
According to the modeling of communication delay, for each state $s\in \cal S$ we formulate the RAILP to maximize the weighted sum of non-zeros in ${\bm \Delta}^{(t)}$ and ${\bm \Lambda}^{(t)}$, i.e, to minimize the average delay of all packets from satellites to GSs.
\begin{equation}
\label{eqn: objRAILP}
\begin{split}
\mathop{\max}\limits_{\bm{X}}\quad & {{\gamma {{\Gamma}^{n \to a}} + (1-\gamma){{\Gamma}^{a \to g}}}} \\
\text{ s.t. }\quad & (1)\textit{-}(4),(6)\textit{-}(8),(18)\textit{-}(27).\\
\end{split}
\end{equation}
Wherein $\gamma$ is a weight factor between the delay from non-anchor satellites to anchor ones and that from anchor satellites to GSs.
Compared with the model in the previous section, this model does not contemplate planning routes for every traffic flow, thus is with far fewer variables.

\section{A Heuristic Based on Maximum Weight Matching}
\label{HGGM}
Although the proposed RAILP method can solve the optimization problem within an acceptable period of time, it may still be time-consuming due to its ILP nature.
When the topology needs rapid reconfiguration, RAILP may become unsuitable for designing the topology. 
In this section, we propose a heuristic based on maximum weight matching (HMWM) to further reduce the complexity. 
Since we resort to algorithms in graph theory, we model the network in each state $s$ as an undirected graph $\cal G(V, E)$, where $\cal V$ is the set of all nodes (satellites and GSs) as stated in Section \ref{sec: system model}, and $\cal E$ is the set of edges including ISLs and GSLs. 
It should be noticed that the edge $e_{i,j}\in \cal E$ stands for the possibility of establishing a link between two nodes, in other words only when $v_i$ and $v_j$ are visible to each other, $e_{i,j}$ can be added to $\cal E$. 
Besides, we denote edges between non-anchor satellites to non-anchor satellites, non-anchor satellites to anchor satellites, anchor satellites to anchor satellites, anchor satellites to ground stations as ${\cal E}^{nn}$, ${\cal E}^{na}$, ${\cal E}^{aa}$, and ${\cal E}^{ag}$, respectively. 
Thus we have ${\cal E} = {\cal E}^{nn} \cup {\cal E}^{na} \cup {\cal E}^{aa} \cup {\cal E}^{ag}$.

In the context of GNSSs, each node is only able to carry one link terminal, making it possible to solve the topology design problem by graph matching algorithms \cite{matching}.  
A \textit{matching} in a graph $\cal G(V, E)$ is a subset $\cal M$ of edges in $\cal G$ such that no two of which meet at a common vertex.
In this paper, we see a non-trivial possibility to obtain time-evolving matching in different time slots by means of sophisticated weighting strategies.
We devise our heuristic drawing on \textit{maximum weight matching}, which produces a matching of maximum total edge weights.
The basic idea of our heuristic is to prioritize edges that are conducive to data flows while taking into consideration the ranging requirement, and obtain the final topology $\bm X$ slot by slot.
The weights of edges $\cal E$ in a particular time slot are determined partly by the traffic difference between the two nodes, and partly by how the edges favor the ranging requirement.  

\subsection{Weights Assignment for Edges and Nodes}
Since we need to optimize the delay of data packets at a given state and satisfy the ranging constraints at the same time, we define the weight of each edge in each time slot as follows: 
\begin{equation}
\label{eqn: edge weight 1}
{w_{i,j,t}} = 
\eta w_{i,j,t}^{{c}} + (1 - \eta )w_{i,j,t}^ {{r}}, 
\quad \forall v_i,v_j \in {{\cal V}}, t \in {\cal T}.
\end{equation}
Wherein, the first part $w_{i,j,t}^{{c}}$ reflects to what degree the communication performance can be optimized if the edge is scheduled, and the second part $w_{i,j,t}^ {{r}}$ is related to the urgency of scheduling this link in the current time slot to meet the ranging requirement, and finally $\eta$ is a weight factor.

In specific, the first part of (\ref{eqn: edge weight 1}), $w_{i,j,t}^{{c}}$, can be given as
\begin{equation}
w_{i,j,t}^{c} = w_{j,i,t}^{c} = 
\begin{cases}
{\rho_{i,t}-\rho_{j,t},} & \forall e_{i,j} \in {\cal{E}}^{na}, \; {v_i \in {\cal V}^n  }\\
{\rho_{i,t},} & \forall e_{i,j} \in {\cal{E}}^{ag},\; {v_i \in {\cal V}^a}\\
{-Q,} & {\forall e_{i,j} \in {\cal{E}}^{nn} \cup {\cal{E}}^{aa}}
\end{cases}
,
\end{equation}
where $\rho_{i,t}$ and $\rho_{j,t}$ are the corresponding weights of nodes $v_i$ and $v_j$. 
If the considered edge is between a non-anchor satellite and an anchor satellite, or between an anchor satellite and a GS, in other words is consistent with the direction of the data flows, this edge is assigned a weight difference of the two end nodes.
We assume all GSs are able to process the received data without congestion, thus the weights $\rho$ for GSs is zero.
Bigger weight difference indicates the edge should be scheduled to alleviate the traffic congestion. 
It should be noted that $w_{i,j,t}^{{c}}$ may appear negative.
Besides, for edges in  ${\cal{E}}^{nn}$ or in ${\cal{E}}^{aa}$, the weights are set to a negative constant value $-Q$.
This is because compared with edges in ${\cal{E}}^{na}$ and  ${\cal{E}}^{ag}$, these edges will not carry any traffic flows and thus we hope such kinds of edges are scheduled as little as possible.
On the other hand, sometimes these edges might also need to be scheduled due to lack of ranging links, so their weights are set to $-Q$ rather than $-\infty$. 

At the beginning of the procedure ($t=1$), we assign the initial traffic profile to each satellite as the node weights.  
\begin{equation}
\label{eqn:node weight 3}
    \rho_{i,1}=f_{i,1}, \quad \forall v_i \in {\cal V}^s.
\end{equation}
After each iteration, we could obtain the matching result $M_t$ corresponding to the $t$-th time slot.
Then on a slot per slot basis, the adjustment of the node weights is subject to the simulated traffic flows as follows:
\begin{equation}
\label{eqn: node weight 1}
\begin{aligned}
    \rho_{i,t+1} = 
    \begin{cases}
        {\max(0,\,\rho_{i,{t}}-C_{ss})+f_{i,t+1},}&{ {e_{i,j} } \in     {\cal{E}}^{na} \cap {{\cal{M}}_{t}}}\\
        {\rho_{i,{t}}+f_{i,t+1},}&{{e_{i,j} } \in {\cal{E}}^{nn} \cap {\cal{M}}_{t}}\\
    \end{cases}
    \\ \forall v_i \in {{\cal V}^n}, 1\le t< T,
    \end{aligned}
\end{equation}
\begin{equation}
\begin{aligned}
\label{}
    \rho_{i,t+1} = 
    \begin{cases}
        {\rho_{i,{t}}+\min(\rho_{j,{t}},\,C_{ss})+f_{i,t+1}}&{e_{i,j}\in {\cal E}^{na} \cap {\cal M}_{t}}\\
        {\max(0, \,\rho_{i,{t}}-C_{sg})+f_{i,t+1},}&{e_{i,j}\in {\cal{E}}^{ag} \cap {{\cal{M}}_{t}}}\\
        {\rho_{i,{t}}+f_{i,t+1},}&{e_{i,j}\in {\cal{E}}^{aa} \cap {\cal{M}}_{t}}\\
    \end{cases}
    \\ \forall v_i \in {{\cal V}^a, 1\le t < T}.
    \end{aligned}
\end{equation}
Recall that $C_{ss}$ and $C_{sg}$ are the capacity (in packets/slot) of ISLs and GSLs respectively.
If an edge $e_{i,j}\in {\cal E}^{na}$ is scheduled, the corresponding non-anchor satellite would send data that do not exceed $C_{ss}$ to the anchor satellite.
The weight of the non-anchor satellite decreases and the weight of the anchor satellite increases accordingly.
The same thing happens if an edge $e_{i,j}\in {\cal E}^{ag}$ is scheduled.
Then for each satellite node, the generated traffic in the current time slot is added as the updated weight.
Besides, for ground stations, we set $\rho_{i,t}=0$ for all time slots $t\in {\cal T}$.

The second part of (\ref{eqn: edge weight 1}), $w_{i,j,t}^ {{r}}$, can be expressed as
\begin{equation}
\label{eqn: ranging weigt}
 w_{i,j,t}^{{r}} = w_{j,i,t}^{{r}} = 
\begin{cases}
      \frac{{\hat w_{i,j,t}^{{r}} + \hat w_{j,i,t}^{{r}}}}{2}, &
   \forall e_{i,j} \in {\cal{E}}^{nn} \cup {\cal{E}}^{na} \cup {\cal{E}}^{aa}\\
   0, &  \forall e_{i,j} \in {\cal{E}}^{ag}
\end{cases}
,
\end{equation}
where $\hat w_{i,j,t}^{{r}}$ is the ranging weight from the perspective of $v_i$ and on the contrary $\hat w_{j,i,t}^{{r}}$ is from the perspective of $v_j$. 
Here, $\hat w_{i,j,t}^{{r}}$ is given as
\begin{equation}
\begin{aligned}
\hat w_{i,j,t}^{{r}} = 
\begin{cases}
{\beta {{\left[ {\frac{{\max (0,\, {L^{\min }} - \sum\limits_{v_j \in {\cal V}^s } {{\ell_{i,j,t - 1}}} )}}{{T - t + 1}}} \right]}^\alpha },}&{ {{\ell}_{i,j,t - 1}} = 0}\\
{0,}&{ {{\ell}_{i,j,t - 1}} = 1}
\end{cases}
\\ \forall v_i,v_j \in {{\cal V}^s}, t \in \cal T. 
\end{aligned}
\end{equation}
Wherein, $\alpha$ and $\beta$ are weight factors and $\ell_{i,j,t-1}$ represents whether satellite nodes $v_i$ and $v_j$ have established a link with each other in the past $t-1$ time slots. 
If $v_i$ has already set up a link with $v_j$ (i.e., ${{\ell}_{i,j,t - 1}} = 1$),  $e_{i,j}$ will weight 0.
And if not,  $\hat w_{i,j,t}^{{r}}$ will take the value related to the ratio of the number of ranging links that still need to be established and the number of time slots left where the topology is not decided.
More specifically, $\sum\nolimits_{v_j \in {\cal V}^s } {{\ell_{i,j,t - 1}}}$ stands for the number of ranging links that $v_i$ has established before the considered $t$-th time slot. 
Recall that $L^{\min}$ is the ranging requirement for all satellite nodes.
If ${  \sum\nolimits_{v_j \in {\cal V}^s } {{\ell_{i,j,t - 1}}} \ge {L^{\min }}}$, the ranging requirement has been met for $v_i$, otherwise more ranging links are supposed to be scheduled.  
In addition, if $\alpha > 1$ the scheduling procedure focuses more on the communication performance when there are enough time slots left (i.e., the ratio is close to 0).
As the ratio becomes bigger,  $\hat w_{i,j,t}^{{r}}$ grows exponentially and the urgency of scheduling such an edge also increases.
It is interesting to note that, when $v_i$ needs more ranging links and $e_{i,j}$ has not been scheduled before (i.e., $ {\ell_{i,j,t - 1}}=0$ and $\hat w_{i,j,t}^{{r}} \neq 0$), but $v_j$ has already established at least $L_{\min}$ ranging links (i.e., $\hat w_{j,i,t}^{{r}} =0$), the ranging weight of edge $e_{i,j}$ should be averaged as in (\ref{eqn: ranging weigt}).
Finally, ${{\ell}_{i,j,t - 1}}$ is given as
\begin{equation}
  \ell_{i,j,t-1} = 
\begin{cases}
    {1,}&{\sum\limits_{t=1}^{t-1} {x}_{i,j,t} \ge 1 }  \\
    {0,}&{\sum\limits_{t=1}^{t-1} {x}_{i,j,t} = 0 }  
\end{cases}
\quad \forall v_i,v_j \in {{\cal V}^s}, 1<t \le T,
\end{equation}
\begin{equation}
\label{eqn: edge weight 2}
    \ell_{i,j,1} = 0, \quad \forall v_i \in {\cal V}^s.
\end{equation}
It can be noted that the definition of $\ell$ here is quite similar to the previous one in (\ref{eqn: ranging links}).
The difference lies in the time span, $\ell$ here considers time slots from 1 to $t-1$, in which the topology has been scheduled.
However $\ell$ in (\ref{eqn: ranging links}) is a variable calculated from topology in all time slots along the state. 

\begin{figure}
    \centering
    \includegraphics[width=0.48\textwidth]{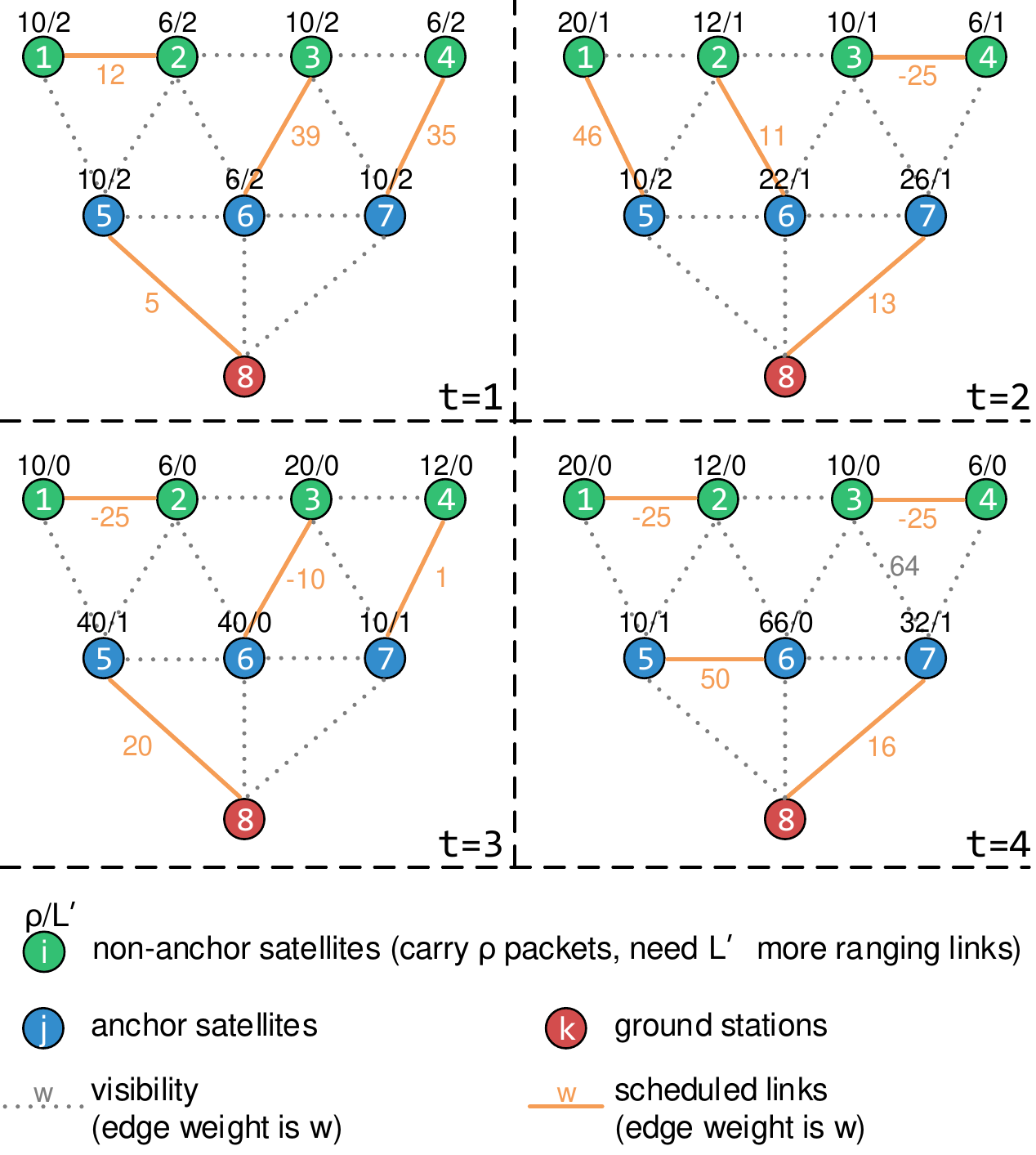}
    \caption{An example of a small-scale HMWM with 8 nodes and 4 time slots: $f^{td}=6$, $f^{sm}=4$, $L^{\min}=2$, $\alpha =2$, $\beta = 300$, $\eta = 0.5$, $Q=50$, satellite nodes with odd indexes should carry short message traffic flows. For simplicity, the weights assigned to edges that are not part of the matching result are omitted. }
    \label{fig: matching}
\end{figure}

\subsection{The Proposed Heuristic and Complexity Analysis}
\label{subsec: procedure}
The specific strategy of HMWM for each state $s\in \cal S$ is shown in Algorithm 1. Let ${\cal M}_t$ represents the matching result in $t$-th slot and $\bm{X} = ({\bm X}_1,\,{\bm X}_2\,...\,{\bm X}_T)$ represent the scheduled topology of $T$ slots. $e_{i,j} \in {\cal M}_t$ means in the matching result, node $v_i$ is matched to $v_j$ in time slot $t$, and thus is equivalent to $x_{i,j,t} = x_{i,j,t} = 1$ in ${\bm{X}}_t$. 
Fig. \ref{fig: matching} gives a simple example of how the weights on nodes and edges determine the scheduled topology in one slot and how the result in turn influences the weights in the next time slot.
In the first time slot, node weights are set according to $f_{i,1}$, and each satellite node needs to establish two ranging links.
For arc $e_{1,2}$, the weight is calculated as $w_{1,2,1} = {\eta}(-Q) + (1-\eta){\beta}\left[ {{L^{\min}}}/(T-t+1)\right]^{\alpha} = 12$.
Note that the weight value is rounded for simplicity.
After the subset ${{\cal M}_1} = \{e_{1,2}, e_{3,6}, e_{4,7},e_{5,8}\}$ with the maximal weight is obtained by matching algorithms, node weights are accordingly updated. 
In the 4-th time slot, $v_3$ has already met the ranging requirement (i.e., with $v_6$ and $v_4$), so $\hat w_{3,7,4}^{{r}} =0$, however $v_7$ still needs one more ranging link, $\hat w_{7,3,4}^{{r}} = \beta \left[ {1}/(T-t+1)\right ]^{\alpha} = 300$. 
The total weight of $e_{3,7}$ is $w_{3,7,4} = 0.5\times (10-32)+0.5\times \left[ (0+300)/{2}\right] = 64$.
It should be noticed that $e_{3,7}$ is with a pretty high weight but it is not in the final matching result ${\cal M}_4$, and thus the ranging requirement $L^{\min} $ is still not satisfied for $v_7$ at the end of the 4-th slot. 
This is because we implement the matching using the perfect matching algorithm, where all nodes of the graph are covered by $\cal M$.
If $e_{3,7}$ is selected, $v_4$ would be a single node and cannot be covered by any means.
\begin{algorithm}
\caption{HMWM}
\label{algo:HMM}
\LinesNumbered 
\KwIn{Network graph $\cal G$ corresponding to the scheduled state, number of time slots $T$, traffic matrix $\bm F$, the required ranging constraint $L^{\min}$, link capacity of ISLs $C_{ss}$, link capacity of GSLs $C_{sg}$, weight factors $\alpha$, $\beta$, and $\eta$.  }
\KwOut{the scheduled topology matrix $\bm X$.}
Initialization: set $t=1$.\\
\For{$t \le T$}
{
Set weight for nodes in $\mathcal G$ according to (31)-(33).\\
Set weight for edges in $\mathcal G$ according to (29), (30), (34)-(37).\\
Compute the maximum matching.\\
According to the matching result ${\cal M}_{t}$, obtain the scheduled topology in the $t$-th time slot, i.e., ${\bm X}_t$.\\
$t=t+1$.
}
Construct $\bm X$ from ${\bm X}_1,\, {\bm X}_2\,...\,{\bm X}_T$. 
\end{algorithm}

We solve the maximal weight matching by an efficient algorithm known as Blossom V \cite{blossomV}, which is based on the well-known Blossom routine \cite{blossom}.
Blossom V runs in time $O(|{\cal E}||{\cal V}|^2)$, where $|\cal E|$ is the number of edges of the graph and $|\cal V|$ is its number of vertices.
Here we assume there are $E$ edges and in our paper $N$ is the number of all nodes. 
For steps 2-8 in Algorithm 1, the loop runs for $T$ times.
The running time of step 3 is $O({N}/{2})$ as in the perfect matching result $N$ nodes are met by exactly ${N}/{2}$ edges.
Besides, step 4 runs in $O(E)$.
For step 5, the running time is $O(EN^2)$, and the time complexity for step 6 is $O(N/2)$.
Finally we can get the conclusion that the running time of the heuristic in Algorithm 1 is $O(T(N/2+E+EN^2+N/2))=O(TEN^2)$.
Note that this time complexity is for one state, thus for $S$ states in the whole scheduled period the time complexity is $O(STEN^2)$.

\section{Performance Evaluation}
\label{sec:Evaluation}

\subsection{Performance in Test Scenario}
\label{subsec:test}
To compare the performance of the proposed methods and the ILP model which is infeasible in practical GNSSs because of complexity, we first perform simulations in a small-scale test scenario with 7 satellites and 1 GS, whose visibility is shown in Fig. \ref{fig: delay} and Fig. \ref{fig: matching}.
Each satellite needs to set up 2 different ISLs, i.e., $L^{\min}$ is set to 2.
Besides, all satellites generate 6 telemetry packets that should be sent to GSs, and half of the satellites would carry additional 4 packets of other services such as GSMC. 
The capacity of the ISLs and GSLs is set to 25 and 50 packets per slot, respectively.
Other parameters are given in Table \ref{tab:parameters}.
As for ILP and RAILP, a powerful toolbox YALMIP \cite{YALMIP} is used to model and solve optimization problems in MATLAB \footnote{Interested readers can access the MATLAB script file of the ILP model and the RAILP model in this paper by: \url{https://github.com/parallelyzb/RAILP.}}.
Finally, we solve the linear programming optimization by one of the off-the-shelf solvers Gurobi. 
All the simulations are performed on a PC with i7-8565U CPU 1.80GHz, 8.00GB RAM.
\begin{table}[htp]
    \caption{Simulation Parameters}
    \label{tab:parameters}
    \centering
    \renewcommand{\arraystretch}{1.1}
    \begin{tabular}{p{2cm}<{\centering}|p{2cm}<{\centering}|p{2cm}<{\centering}}
    \hline
        \multirow{2}{*}{Parameter} & \multicolumn{2}{c}{Value}\\ 
\cline{2-3}
          & Test scenario & Practical GNSS \\ \hline \hline
         $L^{\min}$ & 2 & 6 \\ \hline
         $f^{td}$, $f^{sm}$ & \multicolumn{2}{c}{6, 4 packets}\\ 
\hline
         $C_{ss}$, $C_{sg}$ & \multicolumn{2}{c}{25, 50 packets/slot}\\ \hline
         $B^{\max}$ & \multicolumn{2}{c}{150 packets}\\ \hline
         $\gamma$ & 0.5 & 0.1\\ \hline
         $\eta$, $\alpha$, $\beta$ & 0.4, 2, 700 & 0.7, 2, 500\\ \hline
         Q &  \multicolumn{2}{c}{50}\\ \hline
         $\dot M$, $ M$, $\bar M$, $\tilde M$ &  \multicolumn{2}{c}{25, 50, 25, 25}\\ \hline
         $\{k_1, k_2, k_3 \}$ &------ & {2, 2, 2}\\ \hline
         
    \end{tabular}
\end{table}

Since our work is quite new in GNSSs, we compare the proposed algorithms with a method propsed for remote sensing networks, namely Fair Contact Plan (FCP)\cite{Juan2014} .
Specifically, FCP adopts a matching method which iteratively selects links with maximum accumulated disabled contact time to improve the fairness of the scheduled topology.
Then, the topology results of all slots are randomly exchanged to suit our scenario. 
It is interesting to note that, even FCP is not presented for GNSSs, it can meet the ranging requirement to a certain extent because of its fairness nature. 

\begin{figure}[tp]
    \centering
    \includegraphics[width=3in, height=2.25in]{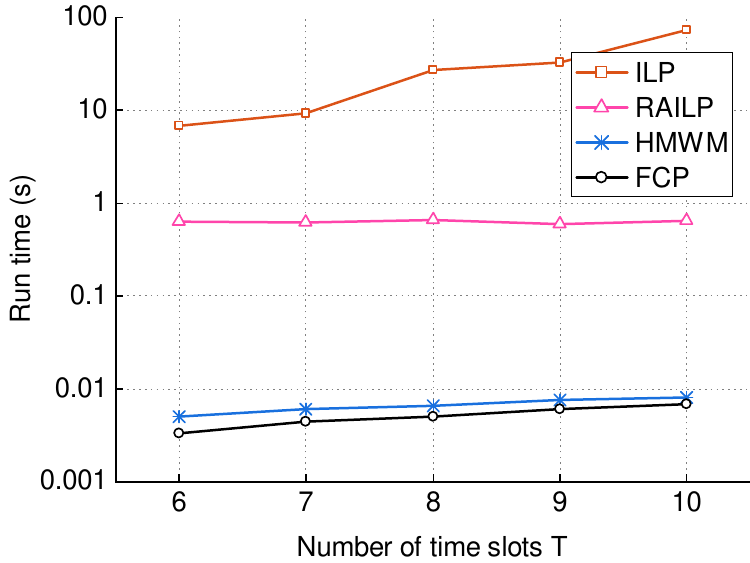}
    \caption{The run time comparison for the test scenario.}
    \label{fig: runtime}
\end{figure}

\begin{figure}[tp]
    \centering
    \includegraphics[width=3in, height=2.25in]{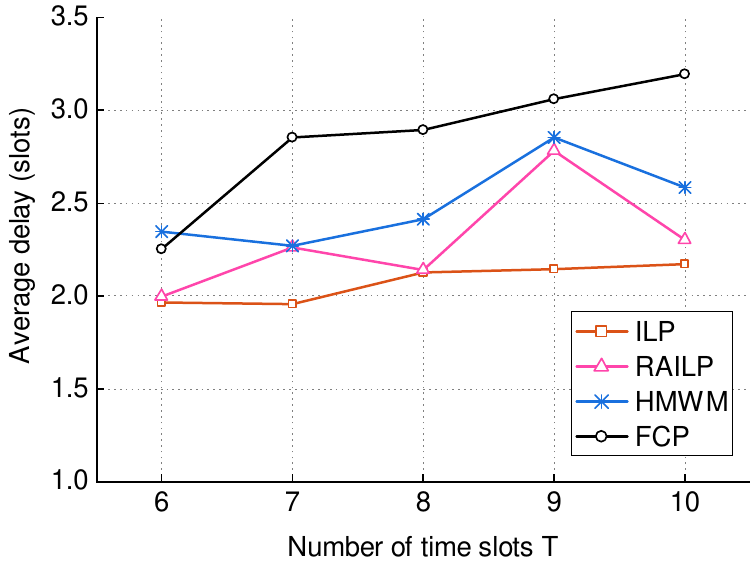}
    \caption{The average delay for the test scenario.}
    \label{fig: testDelay}
\end{figure}

In order to verify the efficiency of our proposed methods, we depict the run time of four different algorithms as shown in Fig. \ref{fig: runtime} with a logarithmic ordinate.
As expected, by decreasing the integer variables our proposed RAILP model can greatly reduce the run time compared with the ILP model.
Due to the Presolve in Gurobi optimizer, the run time of RAILP only increases slightly with the growth of T.
Nevertheless, RAILP is still based on ILP, whose complexity by nature results in relatively long run time in practical GNSSs with more nodes.
On the contrary, HMWM and FCP need remarkably less time to obtain the solutions and also scale well in practical GNSSs due to their heuristics.
It should be noted that in practical GNSSs whose results are given in the next subsection, the average run time of RAILP and HMWM for one state is 11.21s and 0.434s, respectively.
Next, we compare the average delay of data from satellites to GSs as depicted in Fig. \ref{fig: testDelay}.
The statistics are averaged over all packets generated on different satellites in T slots.
It can be seen that the proposed RAILP and HMWM can achieve lower average delay compared with FCP.
Besides, the performance of the RAILP and HMWM is quite close to the optimal value obtained by the ILP model, indicating the suboptimality of the proposed methods.


\subsection{Performance in Practical GNSS}
\label{subsec:results}
Then the simulations are conducted on a practical GNSS with 3 ground stations and 30 satellites including 24 medium earth orbit (MEO) satellites, 3 inclined geosynchronous orbit (IGSO) satellites, and 3 geosynchronous orbit (GEO) satellites \cite{huang2018,Sun2018}.
Table \ref{tab:orbit} gives the detailed orbit parameters of satellites and the location of three GSs.
Each satellite carries one narrow beam directional antenna, whereas each GS could carry several antennas.
We denote the number of antennas on three GSs as $K=\{k_1, k_2, k_3 \}$, and $|K|$ is the total number of antennas of all GSs.
In order to execute the proposed algorithms, each antenna on GSs is regarded as one node. 
The aspect angle of the antennas on MEOs, IGSOs, GEOs and GSs is set to 60°, 45°, 45° and 85°, respectively.
The simulation period is set to 24 hours.
According to \cite{huang2018,Sun2018}, each schedule (i.e., 1 minute) is composed of the topology of 20 time slots lasting 3 seconds. 
To relieve the memory consumption on satellites, the whole period is divided into 288 states (i.e., 5 minutes), and for one state, the topology schedule obtained by topology design algorithms would be reused for 5 times.
The visibility of satellites and GSs is obtained by the contact plan designer plug-in\cite{JuanCPD}.
Other detailed simulation parameters are shown in Table \ref{tab:parameters} unless otherwise specified.
\begin{table}[htp]
\caption{Orbit parameters and GS locations}
\label{tab:orbit}
\centering
\renewcommand{\arraystretch}{1.1}
\begin{tabular}{c|c}
\hline
Satellites and GSs & Description  \\ \hline\hline
MEO    & Walker-$\delta$ 24/3/1, h = 21528km, inc = 55° \\ \hline
IGSO   & h = 35786km, inc = 55°, interval = 120° \\ \hline
GEO    & h= 35786km, lon = (80°, 110.5°, 140°) \\ \hline
Weinan station & (34.2°N, 109.2°E)\\ \hline
Kashi station &  (40.1°N, 79.5°E)\\ \hline
Sanya station &  (18.0°N, 109.3°E)\\ \hline
\end{tabular}
\end{table}

\begin{figure}[tp]
    \centering
    \includegraphics[width=3in, height=2.25in]{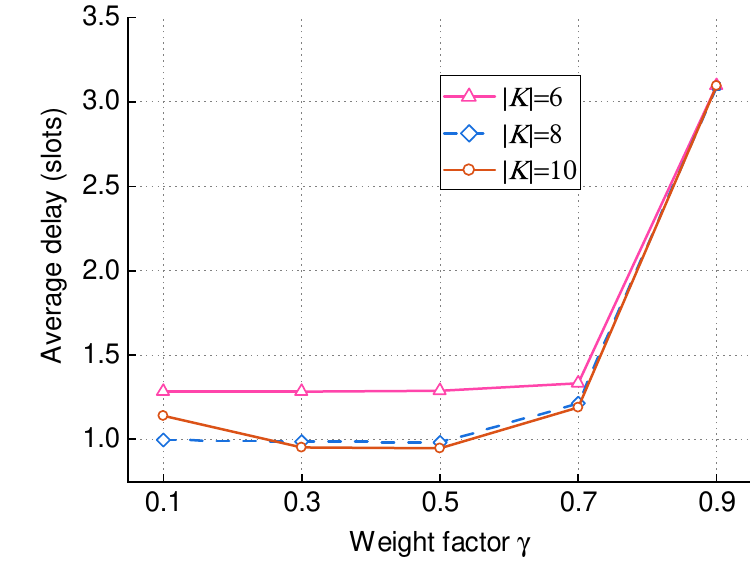}
    \caption{RAILP: average delay v.s. $\gamma$ }
    \label{fig: gamma}
\end{figure}

To figure out the best parameter setting, Fig. \ref{fig: gamma} investigates the impact of $\gamma$ on delay performance with different numbers of GS antennas.
We can observe that when $\gamma$ takes the value of 0.1, 0.3, and 0.5, RAILP provides similar performance in different scenarios.
As $\gamma$ becomes larger to 0.7 and 0.9, there are dramatic increases in the average delay, which is because higher $\gamma$ would lead to emphasis upon delay from non-anchor satellites to anchor satellites and neglect of delay from anchor satellites to GSs.
However, the final delay of non-anchor satellites is actually decided by the two parts, rather than only by the former.
Besides, we can see that the increment of GS antennas would not bring about a linear decrease in delay.
This is because even though anchor satellites have more opportunities to schedule GSLs, they should maintain connection with non-anchor satellites to some extent, otherwise the delay of the non-anchor satellites would become extremely large.

\begin{figure}[tp]
    \centering
    \includegraphics[width=3in, height=2.25in]{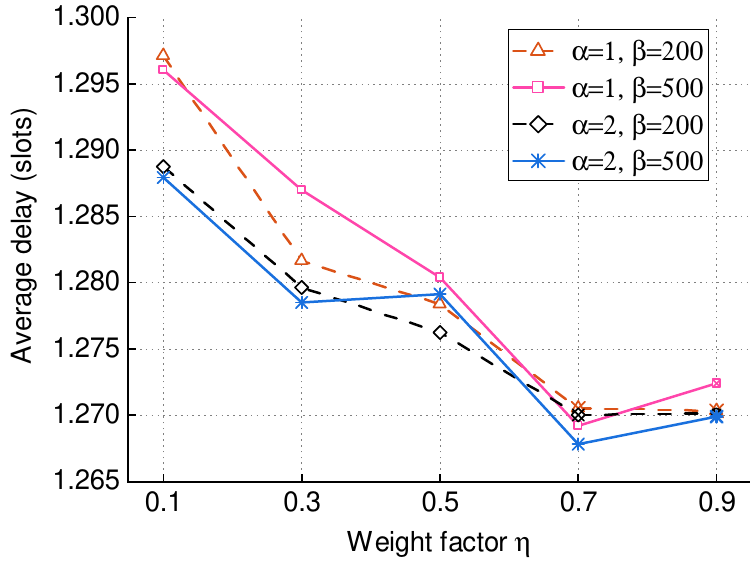}
    \caption{HMWM: average delay v.s. $\eta$ }
    \label{fig: eta}
\end{figure}

Fig. \ref{fig: eta} is plotted to show how the parameters $\alpha$, $\beta$, and $\eta$ affect the performance of topology obtained by HMWM.
As expected, the performance generally rises with the increase of $\eta$, which means to pay more attention to the optimization of communication performance.
However, focusing more on communication performance may result in breach of the ranging requirement $L^{\min}$, which is marked with a cross in the figure.
It can also be seen that HMWM with $\alpha =2$ obtains lower average delay in most cases.
This can be explained by HMWM will focus more on the communication performance when there are enough time slots left if $\alpha > 1$ as stated in Section \ref{HGGM}. 
Finally, $\eta$ together with $\beta$ determine whether the resulting topology could meet the ranging requirement.

\begin{figure}[tp]
    \centering
    \includegraphics[width=3in, height=2.25in]{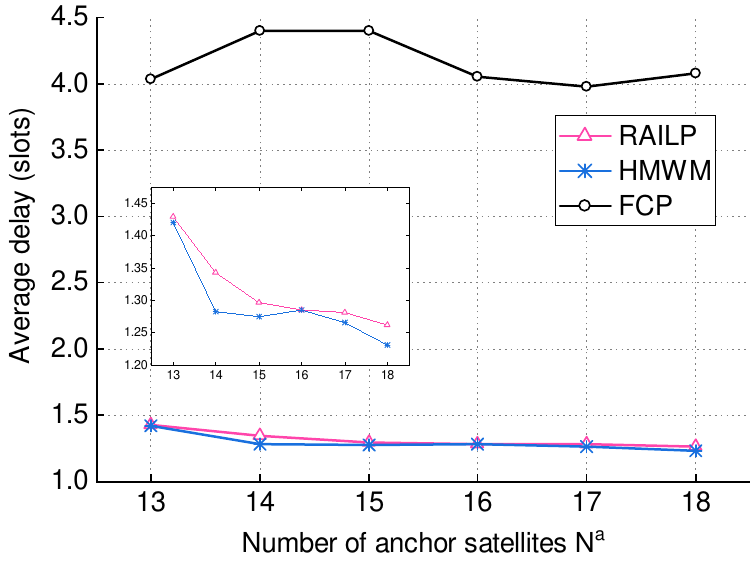}
    \caption{Average delay v.s. $N^a$ }
    \label{fig: anchor}
\end{figure}

\begin{figure}[tp]
    \centering
    \includegraphics[width=3in, height=2.25in]{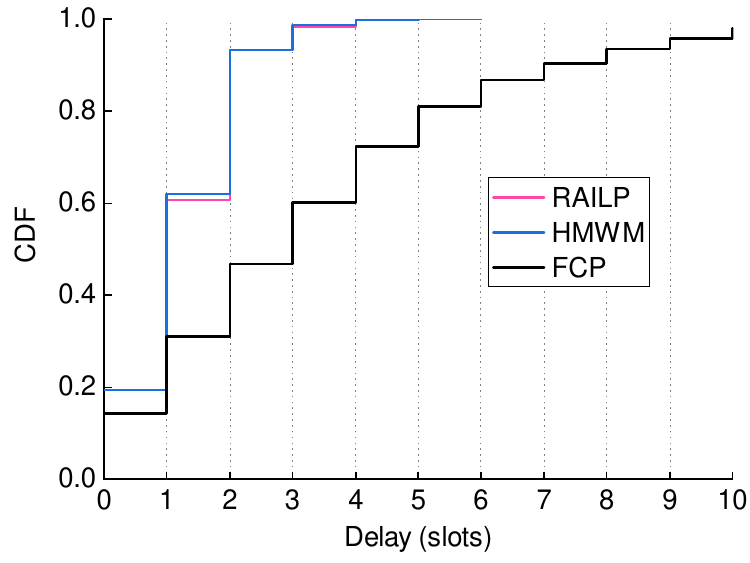}
    \caption{Cumulative Distribution Function (CDF) of delay.}
    \label{fig: CDF}
\end{figure}

Fig. \ref{fig: anchor} shows the performance of the three algorithms with respect to the number of anchor satellites $N^a$. 
Comparing the three algorithms, it can be seen that the proposed RAILP and HMWM can achieve lower average delay regardless of the number of anchor satellites. 
For the proposed two approaches, the average delay decreases with $N^a$ because non-anchor satellites have more opportunities to access anchor satellites and send packets.
However the performance does not improve linearly with the increase of $N^a$, which is because of the finite number of GSs.
Besides, it can be seen that HMWM performs slightly better than RAILP.
This fact can also be seen in Fig. \ref{fig: CDF}, where cumulative distribution of delay are depicted.
The distribution of delay is roughly the same for RAILP and HMWM, and data packets with delay less than or equal to 3 time slots account for 98.3\% and 98.7\%, respectively. 
In addition, FCP provides a maximum delay of more than 10 time slots, whereas the maximum delay of both RAILP and HMWM is 6 slots.
While the proposed algorithms tend to set up links between non-anchor satellites and anchor-satellites or between anchor satellites to GSs, FCP always schedules a fair topology paying no attention to the traffic flows of data transmission, thus resulting in poor performance in delay.

\begin{figure}[tp]
    \centering
    \includegraphics[width=3in, height=2.25in]{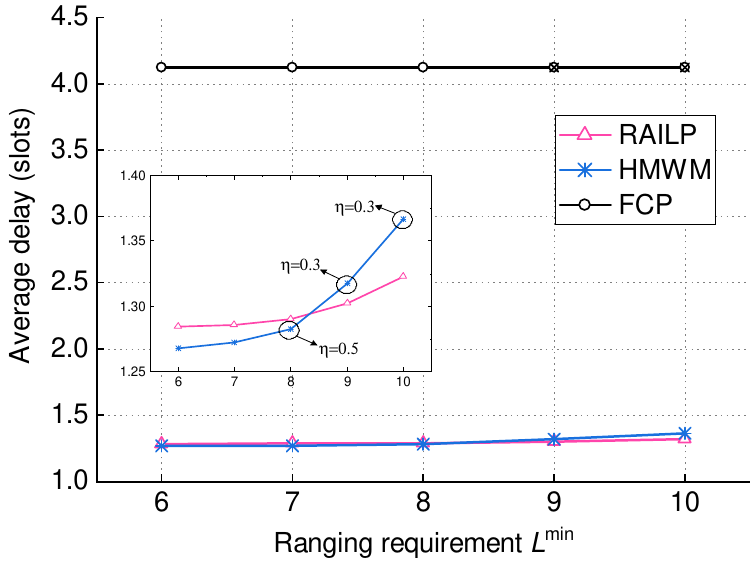}
    \caption{Average delay v.s. $L^{\min}$. }
    \label{fig: Lmin}
\end{figure}

Fig. \ref{fig: Lmin} illustrates how the ranging requirement $L^{\min}$ affects the average delay of GNSS networks.
Since FCP does not consider the ranging constraint, the topology obtained by FCP cannot meet the requirement when $L^{\min}$ gets larger than 8.
When the $L^{\min}$ cannot be satisfied only by scheduling links between non-anchor satellites and anchor satellites, ISLs between non-anchor satellites (i.e., $e \in {\cal{E}}^{nn}$) or between anchor satellites (i.e., $e \in {\cal{E}}^{aa}$) shall be planned.
Therefore with the increase of $L^{\min}$, the average delay of both RAILP and HMWM increases.
It should be noted that for HMWM, $\eta =0.7$ cannot guarantee the satisfaction of $L^{\min}$ when $L^{\min}$ becomes larger.
The corresponding $\eta$ applicable to different simulations is marked on the figure.
In cases where $L^{\min}$ is easily satisfied, HMWM provides slightly better delay than RAILP.
The first reason is because HMWM can focus on optimizing the delay since matching results merely depending on traffic change could partly meet the ranging requirement.  
The second reason is that RAILP is still a sub-optimal algorithm because of the inaccurate modeling of delay from non-anchor satellite to GSs.
However, the superiority of RAILP appears when $L^{\min}$ is 9 or 10.
RAILP could always meet the ranging requirement at the lowest cost and optimize its objection as much as possible because $L^{\min}$ is modeled as a constraint in the linear programming.
On the other hand, for HMWM, the difficulty of determining the boundary where $L^{\min}$ can be appropriately met will penalize the communication metric.

\begin{figure}[tp]
    \centering
    \includegraphics[width=3in, height=2.25in]{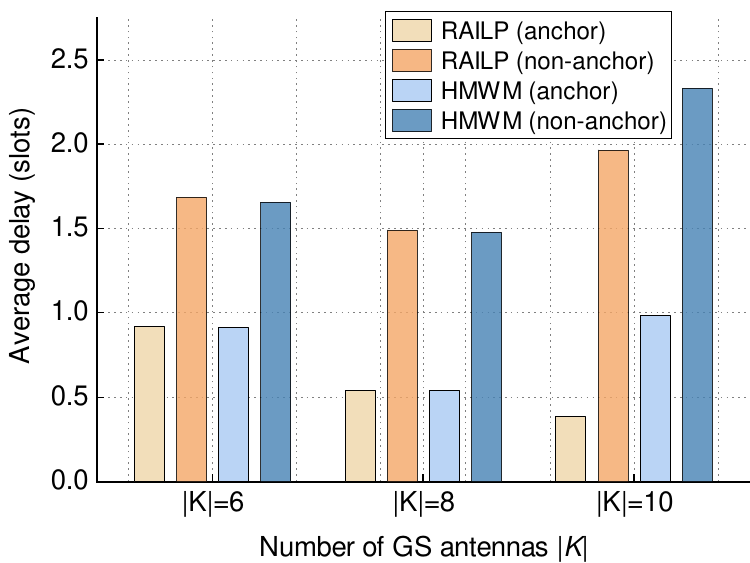}
    \caption{Average delay of anchor and non-anchor satellites. }
    \label{fig: K}
\end{figure}

Finally, Fig. \ref{fig: K} is given to compare the average delay of anchor and non-anchor satellites for different strategies.
Comparing the results of $|K|=6$ and that of $|K|=8$, the average delay of anchor and non-anchor satellites are shortened for both RAILP and HMWM by increasing the number of antennas.
When there are more GS antennas (i.e., $|K|=10$), for RAILP, while the average delay of anchor satellites decreases, the delay of non-anchor satellites increases.
It should be noted that $\gamma =0.1$ is not the best parameter setting when $|K|=10$ as we can see in Fig. \ref{fig: gamma}.
For HMWM, each GS antenna would be scheduled a GSL.
In states with a small number of anchor satellites, most of the anchor satellites are occupied due to the established GSLs.
Thus it becomes hard for the non-anchor satellites to send data to anchor satellites. 
This performance deterioration stems from the perfect matching algorithm as explained in Section \ref{subsec: procedure} and can be mitigated by using non-perfect matching methods.
However, the specific weighting strategies need further study.




\section{Conclusion}
In this paper, we studied the topology design problem in GNSSs where the GSLs adopt the polling scheme like the ISLs do.
Based on the FAS model and the PTDD ranging hierarchy, we formulated the problem as an ILP problem.
To solve the computational infeasibility of ILP in practical GNSSs, we proposed RAILP, a modeling not scheduling routes for each flow, to reduce the number of decision variables.
Aiming at solving the problem more efficiently, we further designed a heuristic named HMWM based on maximum weight matching.
The simulation results proved the suboptimality and efficiency of the proposed algorithms in practical GNSSs.
And compared with RAILP, the proposed HMWM could produce similar performance with respect to average delay but with notably less computational complexity. 
Our work can provide insights on the important but yet underexplored topology design problem in GNSSs where both the ISLs and the GSLs are time-slotted.

Nevertheless, in this paper, we assume only one link, either ISL or GSL, can be set up for one satellite at a given slot.
Deploying laser ISLs on navigation satellites will become increasingly imperative in the foreseeable future \cite{Liu2020}.
By then, each satellite could have multiple simultaneous ISLs and as a result the modeling method of delay in RAILP and the matching algorithm in HMWM will not be applicable any more.
Thus we leave as future work the exploration of corresponding topology design algorithms in GNSSs with laser ISLs.
\bibliographystyle{IEEEtran}
\bibliography{bib}

\begin{thebibliography}{10}
\providecommand{\url}[1]{#1}
\csname url@samestyle\endcsname
\providecommand{\newblock}{\relax}
\providecommand{\bibinfo}[2]{#2}
\providecommand{\BIBentrySTDinterwordspacing}{\spaceskip=0pt\relax}
\providecommand{\BIBentryALTinterwordstretchfactor}{4}
\providecommand{\BIBentryALTinterwordspacing}{\spaceskip=\fontdimen2\font plus
\BIBentryALTinterwordstretchfactor\fontdimen3\font minus
  \fontdimen4\font\relax}
\providecommand{\BIBforeignlanguage}[2]{{%
\expandafter\ifx\csname l@#1\endcsname\relax
\typeout{** WARNING: IEEEtran.bst: No hyphenation pattern has been}%
\typeout{** loaded for the language `#1'. Using the pattern for}%
\typeout{** the default language instead.}%
\else
\language=\csname l@#1\endcsname
\fi
#2}}
\providecommand{\BIBdecl}{\relax}
\BIBdecl

\bibitem{PNT}
F.~Pereira and D.~Selva, ``{Tradespace analysis of GNSS Space Segment
  Architectures},'' \emph{{IEEE Trans. Aerosp. Electron. Syst.}}, vol.~{57},
  no.~{1}, pp. {155--174}, {Feb.} {2021}.

\bibitem{yang2017}
D.~Yang, J.~Yang, and P.~Xu, ``{Timeslot scheduling of inter-satellite links
  based on a system of a narrow beam with time division},'' \emph{{GPS
  Solut.}}, vol.~{21}, no.~{3}, pp. {999--1011}, {Jul.} {2017}.

\bibitem{huang2018}
J.~Huang, W.~Liu, Y.~Su, and F.~Wang, ``{Cascade optimization design of
  inter-satellite link enhanced with adaptability in future GNSS satellite
  networks},'' \emph{{GPS Solut.}}, vol.~{22}, no.~{2}, {Apr.} {2018}.

\bibitem{Sun2018}
L.~Sun, Y.~Wang, W.~Huang, J.~Yang, Y.~Zhou, and D.~Yang, ``{Inter-satellite
  communication and ranging link assignment for navigation satellite
  systems},'' \emph{{GPS Solut.}}, vol.~{22}, no.~{2}, {Apr.} {2018}.

\bibitem{gps}
O.~Luba, L.~Boyd, A.~Gower, and J.~Crum, ``{GPS III system operations
  concepts},'' \emph{{IEEE Aerosp. Electron. Syst. Mag.}}, vol.~{20}, no.~{1},
  pp. {10--18}, {Jan.} {2005}.

\bibitem{Galilo}
M.~{S´anchez}, J.~{Pulido}, F.~{Amarillo}, and J.~{Gerner}, ``{The ESA GNSS+
  project. Inter-satellite ranging and communication links in the frame of the
  GNSS infrastructure evolutions},'' in \emph{{Proc. 21st Int. Tech. Meeting
  Satell. Div. Inst. Navigat.}}, Savannah, GA, USA, {2008}, pp. 2538 -- 2546.

\bibitem{beidou}
J.~Liu, T.~Geng, and Q.~Zhao, ``{Enhancing precise orbit determination of
  compass with inter-satellite observations},'' \emph{{Surv. Rev.}}, vol.~{43},
  no. {322}, pp. {333--342}, {Sep.} {2011}.

\bibitem{yang2018}
G.~{Yang}, R.~{Wang}, A.~{Sabbagh}, K.~{Zhao}, and X.~{Zhang}, ``Modeling
  optimal retransmission timeout interval for bundle protocol,'' \emph{IEEE
  Trans. Aerosp. Electron. Syst.}, vol.~54, no.~5, pp. 2493--2508, 2018.

\bibitem{yangdaoning2017}
D.~Yang, J.~Yang, G.~Li, Y.~Zhou, and C.~Tang, ``{Globalization highlight:
  orbit determination using BeiDou inter-satellite ranging measurements},''
  \emph{{GPS Solut.}}, vol.~{21}, no.~{3}, pp. {1395--1404}, {Jul.} {2017}.

\bibitem{yang2019}
Y.~Yang, W.~Gao, S.~Guo, Y.~Mao, and Y.~Yang, ``{Introduction to BeiDou-3
  navigation satellite system},'' \emph{{Navigation}}, vol.~{66}, no.~{1}, pp.
  {7--18}, {Jan.} {2019}.

\bibitem{LI2020}
G.~Li, S.~Guo, J.~Lv, K.~Zhao, and Z.~He, ``Introduction to global short
  message communication service of beidou-3 navigation satellite system,''
  \emph{Adv. Space Res.}, vol.~{67}, no.~{5}, pp. {1701--1708}, Mar. 2021.

\bibitem{JuanSurvey}
J.~A. {Fraire} and J.~M. {Finochietto}, ``Design challenges in contact plans
  for disruption-tolerant satellite networks,'' \emph{IEEE Commun. Mag.},
  vol.~53, no.~5, pp. 163--169, May 2015.

\bibitem{Juan2014}
J.~A. Fraire, P.~G. Madoery, and J.~M. Finochietto, ``On the design and
  analysis of fair contact plans in predictable delay-tolerant networks,''
  \emph{IEEE Sensors J.}, vol.~14, no.~11, pp. 3874--3882, Nov. 2014.

\bibitem{Yan2015}
H.~Yan, Q.~Zhang, Y.~Sun, and J.~Guo, ``Contact plan design for navigation
  satellite network based on simulated annealing,'' in \emph{Proc. IEEE Int.
  Conf. Commun. Softw. Netw.}, Jun. 2015, pp. 12--16.

\bibitem{Hou2018}
Z.~Hou, X.~Yi, Y.~Zhao, C.~Li, and Y.~Xie, ``{Contact Plan Design for
  Navigation Satellite Network Based on Maximum Matching},'' in \emph{Proc. ACM
  Int. Conf. Proc. Ser.}, Las Vegas, NV, USA, {Aug.} {2018}, pp. 1--6.

\bibitem{YanWCL}
Z.~{Yan}, G.~{Gu}, K.~{Zhao}, Q.~{Wang}, G.~{Li}, X.~{Nie}, H.~{Yang}, and
  S.~{Du}, ``Integer linear programming based topology design for gnsss with
  inter-satellite links,'' \emph{IEEE Wireless Commun. Lett.}, vol.~10, no.~2,
  pp. 268--290, Feb. 2021.

\bibitem{YanTAES}
Z.~{Yan}, J.~A. {Fraire}, K.~{Zhao}, H.~{Yan}, P.~G. {Madoery}, W.~{Li}, and
  H.~{Yang}, ``Distributed contact plan design for gnsss,'' \emph{IEEE Trans.
  Aerosp. Electron. Syst.}, vol.~56, no.~1, pp. 660--672, Feb. 2020.

\bibitem{Ren2019}
X.~Ren, Y.~Yang, J.~Zhu, and T.~Xu, ``Comparing satellite orbit determination
  by batch processing and extended kalman filtering using inter-satellite link
  measurements of the next-generation beidou satellites,'' \emph{GPS Solut.},
  vol.~23, no.~1, 2019.

\bibitem{Bai2020}
Y.~{Bai}, Y.~{Guo}, X.~{Wang}, and X.~{Lu}, ``Satellite-ground two-way
  measuring method and performance evaluation of bds-3 inter-satellite link
  system,'' \emph{IEEE Access}, vol.~8, pp. 157\,530--157\,540, 2020.

\bibitem{Chang1998}
H.~Chang, B.~Kim, C.~Lee, S.~Min, Y.~Choi, H.~Yang, D.~Kim, and C.~Kim,
  ``{FSA-based link assignment and routing in low-earth orbit satellite
  networks},'' \emph{{IEEE Trans. Veh. Technol.}}, vol.~{47}, no.~{3}, pp.
  {1037--1048}, {Aug.} {1998}.

\bibitem{Juan2016}
J.~A. Fraire, P.~G. Madoery, and J.~M. Finochietto, ``{Traffic-aware contact
  plan design for disruption-tolerant space sensor networks},'' \emph{{Ad Hoc
  Netw.}}, vol.~{47}, pp. {41--52}, {Sep.} {2016}.

\bibitem{Zhou2017}
D.~{Zhou}, M.~{Sheng}, X.~{Wang}, C.~{Xu}, R.~{Liu}, and J.~{Li}, ``Mission
  aware contact plan design in resource-limited small satellite networks,''
  \emph{IEEE Trans. Commun.}, vol.~65, no.~6, pp. 2451--2466, {Jun.} 2017.

\bibitem{Zhou2019}
D.~{Zhou}, M.~{Sheng}, B.~{Li}, J.~{Li}, and Z.~{Han}, ``Distributionally
  robust planning for data delivery in distributed satellite cluster network,''
  \emph{IEEE Trans. Wirel. Commun.}, vol.~18, no.~7, pp. 3642--3657, 2019.

\bibitem{Bhattacherjee2019}
D.~Bhattacherjee and A.~Singla, ``Network topology design at 27,000 km/hour,''
  in \emph{Proc. Int. Conf. Emerg. Netw. Exp. Technol.}, NY, USA, 2019, p.
  341–354.

\bibitem{Bertsekas1997}
D.~P. Bertsekas, ``Nonlinear programming.'' \emph{J. Oper. Res. Soc.}, vol.~48,
  no.~3, pp. 334--334, Mar. 1997.

\bibitem{Gao2012}
G.~X. {Gao} and P.~{Enge}, ``How many gnss satellites are too many?''
  \emph{IEEE Trans. Aerosp. Electron. Syst.}, vol.~48, no.~4, pp. 2865--2874,
  Oct. 2012.

\bibitem{matching}
L.~{Lovász} and M.~{Plummer}, \emph{Matching theory}.\hskip 1em plus 0.5em
  minus 0.4em\relax American Mathematical Soc., 2009, vol. 367.

\bibitem{blossomV}
V.~{Kolmogorov}, ``Blossom v: A new implementation of a minimum cost perfect
  matching algorithm,'' \emph{Math. Program. Comput.}, no.~1, pp. 43--67, Jul.
  2009.

\bibitem{blossom}
J.~{Edmonds}, ``Path, trees, and flowers,'' \emph{Can. J. Math.}, pp. 449--467,
  1965.

\bibitem{YALMIP}
J.~{Löfberg}, ``Yalmip: A toolbox for modeling and optimization in matlab,''
  in \emph{Proc. IEEE Int. Symp. Comput. Aid Control Syst. Des.}, 2004, pp.
  284--289.

\bibitem{JuanCPD}
J.~A. {Fraire}, ``Introducing contact plan designer: A planning tool for
  dtn-based space-terrestrial networks,'' in \emph{Proc. IEEE Int. Conf. Space
  Mission Challenges Inf. Technol.}, 2017, pp. 124--127.

\bibitem{Liu2020}
S.~Liu, J.~Yang, X.~Guo, and L.~Sun, ``Inter-satellite link assignment for the
  laser/radio hybrid network in navigation satellite systems,'' \emph{GPS
  Solut.}, vol.~24, no.~2, p.~49, 2020.

\end{thebibliography}

\end{document}